\documentclass[11pt]{article}

\usepackage[margin=2.5cm]{geometry}
\usepackage{amsthm}
\usepackage{amsmath}
\usepackage{amsfonts}
\usepackage{graphicx}
\usepackage{enumerate}

\newtheorem{thm}{Theorem}[section]
\theoremstyle{definition}
\newtheorem{dfn}{Definition}[section]
\theoremstyle{remark}

\theoremstyle{plain}
\newtheorem{lem}[thm]{Lemma}
\newtheorem{prop}{Proposition}[section]

\newtheorem{ass}{Assumption}[section]

\newcommand{\calh}{\mathcal{H}}

\newcommand{\cals}{\mathcal{S}}

\newcommand{\calt}{\mathcal{T}}

\newcommand{\tp}{\otimes}
\newcommand{\tg}{\textbf{tg}}
\newcommand{\T}{\textbf{T}}
\newcommand{\R}{\textbf{R}}

\newcommand{\N}{\textbf{N}}
\newcommand{\C}{\textbf{C}}

\begin{document}
\title{Some convolution products in Quantum Field Theory}
\author{H.M. Ratsimbarison\\
       Institut @HEP-Mad, Antananarivo}
\date{May 2006}

\maketitle

\begin{abstract} This paper aims to show constructions of scale dependence and interaction on some probabilistic models which may be revelant for renormalization theory in Quantum Field Theory. We begin with a review of the convolution product's use in the Kreimer-Connes formalism of perturbative renormalization. We show that the Wilson effective action can be obtained from a convolution product propriety of regularized Gaussian measures on the space of fields. Then, we propose a natural C*-algebraic framework for scale dependent field theories which may enhance the conceptual approach to renormalization theory. In the same spirit, we introduce a probabilistic construction of interacting theories for simple models and apply it for quantum field theory by defining a partition function in this setting. 
\end{abstract}

\section{Introduction}

Quantum field theory (QFT) is a physical theory which combines field theory, already used in Classical mechanics to describe Electromagnetism, and quantum mechanical principles, believed to govern the behavior of microscopic systems. The most computable approach of QFT is the path integral formalism. In this setting, free QFT is assumed to be a formal Gaussian probability law on a space of fields and interacting QFT extends the formal Gaussian law by adding an additional term, the interacting term, to the quadratic term of the free QFT. \\[9pt]
Besides being mathematically formal, the path integral formalism of QFT are spoiled by divergence problems because most of relevant quantities calculated in are divergent. Theses divergences is essentially classified into ultra-violet (UV) divergences and infra-red (IR) ones. By definition, UV divergences appears when one integrates over a space of fields with arbitrarily high momentum modes, or equivalently in the integration over fields localized on the same point of spacetime \footnote{In this paper, we will treat essentially UV divergences.}\cite{acdk99}. The elimination of these divergences in QFT leads to the renormalization theory, initiated in the forties by physicists such as Richard Feynman, Sin-Itiro Tomonaga, and Julian Schwinger. Later, in the seventies, it permits to understand the dependence of QFT with the energy scale and lead to important notions such as \emph{fundamental}, \emph{effective theories}, and \emph{decoupling}. \\[9pt]
Among the renormalization techniques, those of Bogoliubov-Parasiuk-Hepp-Zimmermann (BPHZ) consists in eliminating divergences by adding counter-terms to amplitudes; the counter-terms are constructed in a recursive way in the presence of subdivergent amplitudes. There is also the so-called Dimensional Regularization with minimal subtraction which consists in perturbating the spacetime dimension (Dim Reg) of an integration over the spacetime (Dim Reg-MS). Later, Dirk Kreimer have shown that the BP preparation allows to define a Hopf structure on some set of Feynman diagrams. More later, Alain Connes and Dirk Kreimer introduced a more natural picture for understanding the relations which give counter-term and renormalized amplitude: it is the Birkhoff decomposition. In simplified terms, counter-term and renormalized amplitude are the duals (in the sense of Gelfand-Naimark) of the Birkhoff decomposition components of a loop \cite{acdk99,acdk00}.\\[10pt]
In this paper, we will focus on two convolution products occuring in the renormalization theory and the path integral formalism. The first convolution product will allow us to build the renormalized Feynman amplitude from the nonrenormalized one and a counterterm thanks to the Kreimer's coproduct. Next, the Wilson effective action can be constructed when the regularized free measures are equipped with a convolution product induced by the linear decomposition of fields following their Fourier modes. A brief digression to the probabilistic nature of the Legendre effective action will rienforce the natural presence of probability theory in QFT. Knowing the importance of scale dependence of physical theories, the next section gives a mathematical framework which can derive naturally this physical concept. We will conclude with a probabilist construction of interacting theories for some simple models and a definition of partition function in this setting.

\section{Partition functions in QFT}

In this paper, the path integral formalism of QFT will deal with partition functions defined by means of:
\begin{itemize}
	\item a space of complex valued fields denoted by Fields $\ni \phi$; (classical) fields allows to modelize matters and interactions in Particle Physics, such as the spinor field obeying the Dirac equation and the electromagnetic field governed by the Maxwell's equations.
	\item a formal measure D$\phi$ on Fields such that
\begin{eqnarray}
	D\phi|_{Fields_1\oplus Fields_2} = D\phi|_{Fields_1} \ast D\phi|_{Fields_2} , \quad \textrm{for } \; Fields_1\oplus Fields_2 \, \subseteq \, Fields.
	\label{mesfico}
\end{eqnarray}
where $\ast$ denotes the convolution of measures induced by the addition operation on Fields (see Appendix ).\\
The uses of this formal measure is essentially motivated by the belief that path integral formalism is a generalization of the quasi-Gaussian integral theory, as sustained by applications of Wick theorem, the Stationary phase approximation, etc, in Quantum Field Theory.
  \item an action S defined on Fields; the notion of action takes its origines in the Least action Principle in Classical Mechanics. Among the possible fields of a given physical system, the 'real' field corresponds to those which extremizes a real-valued function on Fields called \emph{action}.\\
  In general, an action S takes the following form:
  \begin{eqnarray*}
	S := S_{free} + S_{int}, \quad S_{int}(\phi) := \sum_{m\geq 1} \frac{g_m}{m!}\,Q_m(\phi^{\otimes m})
\end{eqnarray*}
where the free term S$_{free}$ is a definite positive quadratic form on Fields\footnote{The linear structure on $\C$ induces a natural one on Fields.}, $Q_m$ a symmetric form of order m on Fields, and $g_m$'s formal parameters used to separate the contributions from $Q_m$'s.
\end{itemize}
Now, let us give the definition of partition function in the path integral formalism.
\begin{dfn} Let S - J, J $\in$ Fields$^*$, be an action with an external source J on Fields, then the \emph{normalized (Euclidian) partition function with source} Z$_{S-J}$ associated to S - J is defined by:
\begin{eqnarray*}
	Z_{S-J} := \int_{Fields} D\phi \; e^{-(S - J)(\phi)} = \int_{Fields} D\phi \; e^{-S(\phi) + (J,\phi)},
\end{eqnarray*}
with the normalization condition
\begin{eqnarray}
	Z_{S_{free}} := \int_{Fields} D\phi\; e^{-S_{free}(\phi)} = 1. 
	\label{norcon}
\end{eqnarray}
\end{dfn}
\textbf{Remarks}:
\begin{itemize}
	\item The normalization condition (\ref{norcon}) means that S$_{free}$ defines a probability measure $\mu_{free}$ on Fields by:
\begin{eqnarray*}
	\mu_{free}(f) := \int_{Fields} D\phi\; e^{-S_{free}(\phi)}f(\phi), 
\end{eqnarray*}
for any formally integrable $\C$-valued function f on Fields.
	\item From the above remark, Z$_S$ is then the expectation value of e$^{-S_{int}}$ by $\mu_{free}$.  
\end{itemize}

Important quantities useful in calculation of cross section and S-matrix are correlation functions whose definition is required the introduction of a spacetime in QFT. So let M be a vector space manifold (a spacetime), Fields = $\cals(M)$ the set of all smooth rapidly decreasing $\C$-valued functions on M (the Schwartz space on M), and $\psi: \cals(M) \rightarrow End(\calh)$ a quantization map to the Hilbert space $\calh$, then the n-point correlation function G$^{(N)}$ defined on M$^N$ is given by:
\begin{eqnarray*}
	G^{(N)}(x_1,...,x_N) &:=& \left\langle \Omega|\calt \left[\psi(\delta_{x_{1}})\cdot...\cdot \psi(\delta_{x_{N}})\right]|\Omega \right\rangle,\quad x_1,...,x_N \in M, \: N\in \N^*, \\
	&=& \frac{\int_{\cals(M)} D\phi\; \phi(x_1)...\phi(x_N)\,e^{-S(\phi)}}{\int_{\cals(M)} D\phi \;e^{-S(\phi)}} , \quad \textrm{(Feynman-Kac formula)}
\end{eqnarray*}
where $\Omega$ is the vacuum state of ($\calh$,S), $\delta_x$ the Dirac distribution localized on x$\in$M ($\delta_x$($\phi$) := $\phi$(x), f$\in\cals$(M)), and the time ordering operation $\calt$ arranges operators $\psi(\delta_{x_i})$ in order of increasing time from right to left.\\[10pt]
Thanks to works of Richard Feynman on QFT, there exists a perturbative way to express (\emph{not to evaluate} \footnote{The pertubative method does not gives an evaluation because the results are often divergent.}) Z$_S$ and G$^{(N)}$ by the use of the so-called Feynman graphs.
\par
A \textsl{graph} is a collection of vertices and edges; each edge 'contracts' two vertices and the number of edges incident on a vertex is called its \textsl{valence}. A \textsl{Feynman graph} is a graph with 'external' (or labeled) vertices, which are all univalent, and 'internal' (or bar-labeled) ones, of different valences. We denote by FG(N,$\vec{n}$) the set of equivalence classes of all graphs which have N external vertices, and $n_m$ m-valent internal vertices , $m \geq 1 , n_m \in \vec{n}$.
\par 
To get a pertubative expression of Z$_S$ and G$^{(N)}$, we use Feynman graphs on FG(0) := $\cup_{\vec{n}}$FG(0,$\vec{n}$) and FG(N,$\vec{n}$) respectively, and we associate to every Feynman graph $\Gamma$ a complex number called \emph{Feynman amplitude} $F_\Gamma $ following the so-called \emph{Feynman rules}:\\
(1) place at each vertex some symmetric tensor in V;
\begin{itemize}
	\item At external vertex x place form $\delta_x$,
	\item At each m-valent internal vertex place tensor $Q_m$ , 
\end{itemize}
(2) and contract the tensors along the edges with the inverse form of $S_{free}$,\\[10pt]
and we obtain \cite{paet02}:
\begin{eqnarray*}
	&& G^{(N)}(x_1,...,x_N) = \sum_{\Gamma \in FG(N)} \omega_{\Gamma}\,F(\Gamma), \quad \omega_{\Gamma} := \frac{\prod_i {g_i^{n_i}}}{|Aut(\Gamma)|} \quad \textrm{ for } \Gamma \in FG(N,n_i),\: N\in \N, \\
	&& Z_{S-J} = \sum_{N\in \N} \frac{1}{N!}\,\left\langle J^N,G^{(N)}\right\rangle, \quad \forall\, J\in \cals(M)^*.
\end{eqnarray*}
where $\left\langle ,\right\rangle$ is the tensor pairing. \\
These perturbative calculations remain formal because most of the Feynman amplitudes are divergent. In order to obtain corresponding numerical values, we need to eliminate these divergences. Actually, it can be done by the so-called Connes-Kreimer perturbative renormalization which is a mathematical explanation of the BPHZ renormalization.

\section{Connes-Kreimer renormalization in QFT}

The Connes-Kreimer perturbative renormalization in QFT consists in defining a Hopf algebra structure on the vector space H generated by 1PI graphs and their disjoint unions. The main structure on H is the coproduct $\Delta$ which allows to decompose a graph $\Gamma$ into its divergent subgraph $\gamma$ and its corresponding quotient $\Gamma/\gamma$ \cite{acdk99}, and we have:
\begin{eqnarray*}
	\Delta(\Gamma) = \sum_{\gamma \subseteq \Gamma}\gamma \otimes \Gamma /\gamma = \Gamma \otimes 1 + 1 \otimes \Gamma + \sum_{\gamma \subset \Gamma}\gamma \otimes \Gamma /\gamma , \quad \Gamma \in H.
\end{eqnarray*}
The coproduct on H encodes the combinatorial structure of the Bogoliubov-Parasiuk (BP) preparation which allows to eliminate (sub)divergences of Feynman diagrams.
\par
Most precisely, for a divergent connected graph $\Gamma$,\\
the prepared graph P($\Gamma$) is given by:
\begin{eqnarray*}
	P(\Gamma) = F(\Gamma) + \sum_{\gamma \subset \Gamma}C(\gamma)F(\Gamma/\gamma), \quad \Gamma \in H,
\end{eqnarray*}
where the sum is on all subdivergent graphs, (F is the Feynman amplitude map)\\
the counter-term by:
\begin{eqnarray}
	C(\Gamma) = -T(P(\Gamma)) = -T(F(\Gamma) + \sum_{\gamma \subset \Gamma}C(\gamma)F(\Gamma/\gamma)), \quad \Gamma \in H,
	\label{ct}
\end{eqnarray}
and the renormalized amplitude by:
\begin{eqnarray}
	R(\Gamma) = P(\Gamma) + C(\Gamma), \quad \Gamma \in H,
	\label{reap}
\end{eqnarray}
where T is the operation defining the counter-term for a graph without subdivergence. The renormalization procedure of the BPHZ method is contained in the T-operation\footnote{The operation T amounts to the projection on pole part of a Laurent series in the MS scheme.}.\\
When we define the convolution product $\ast$ associated to $\Delta$ by:
\begin{eqnarray*}
	f \ast g = m_\C \circ f\otimes g \circ \Delta, \quad \textrm{for} \; f,g\in Map(\calh,\C), 
\end{eqnarray*}
then we easily see that R is given by the convolution product of C by F, i.e.	R = C$\ast$F \cite{hera061}.\\
Therefore, one can define the renormalized partition function Z$_{S,ren}$ by:
\begin{eqnarray*}
	Z_{S,ren} := \sum_{\Gamma \in FG(0)} \omega_{\Gamma}\,R(\Gamma) = \sum_{\Gamma \in FG(0)} \omega_{\Gamma}\,C\ast F(\Gamma).
\end{eqnarray*}
Now, we claim that the renormalization theory in QFT consists in modifying Z$_S$ into Z$_{S,ren}$ = 1. 

\begin{ass} Z$_{S,ren}$ = 1.
\label{as1} 
\end{ass}
One argument in favor of this assumption is the quantum-mechanical definition of correlators in QFT. Indeed, it is natural to \emph{define} the renormalized partition function by
\begin{eqnarray*}
	Z_{S,ren} := G^{(0)} := \left\langle \Omega|\Omega \right\rangle.
\end{eqnarray*}
Another reason in favor of Assumption \ref{as1} is that by construction, Z$_{S,ren}$	should be finite and a well-chosen redefinition (by multiplicative factor) of the counter-term C would normalize the renormalized partition function. \\
Following the path integral formalism, it is then natural to define a QFT as a probability law on a space of fields. Two important constructions in non perturbative renormalization theory which we will develop in the next section, confirm largely the probabilistic nature of QFT.

\section{Effective theory in QFT}

Effective theory is an alternative way for renormalizing QFT in a more rigorous manner than in the perturbative treatment. In general, to construct an effective theory means to construct an effective action of the initial bare theory \footnote{However, we must be careful because there are many terms about effective actions which may create confusions}.\\
Let us begin with the derivation of effective theory concepts from renormalization theory. To remove ultra-violet divergences on correlation functions, we can:
\begin{enumerate}
	\item introduce a cutoff (an energy upperbound of field's momentum) in the Feynman (integral) amplitude to eliminate divergences: it is the \emph{regularization procedure}. In some cases, one can achieve this on all correlators by regularizing only the propagator of the theory;
	\item re-express bare physical paramaters in terms of renormalized paramaters in such a manner that the cutoff appears in their expressions; expressions of the renormalized couplings contain also a scale taken from experiment \footnote{The dependence of physical parameters with their scale of observation is described by the renormalization group (RG) equation of the theory.}. This step is the \emph{renormalization procedure};
	\item remove the cutoff to infinity. 
\end{enumerate}
As said above, the cutoff can be interpreted as a momentum \emph{scale} which gives an (upperbound) characteristic value of momentums defined in the physical theory. There is 2 essential scales: the scale introduced in the regularization procedure which defines the \emph{bare} theory, and those required by the renormalization procedure which defines the \emph{effective} theory.\\[10pt]
For a given bare theory (L,$\Lambda$) at scale $\Lambda$, the goal is to construct a theory R$_{\Lambda \mu}$(L,$\Lambda$) at scale $\mu$ equivalent to (L,$\Lambda$) in the following sense \cite{dagr196}: 
\begin{eqnarray*}
	G^{(n)}(R_{\Lambda \mu}(L,\Lambda)) = G^{(n)}(L,\Lambda) \quad \forall\,n\in \N.
\end{eqnarray*}
In this case, R$_{\Lambda \mu}$(L,$\Lambda$) is called an \emph{effective theory} at scale $\mu$ of (L,$\Lambda$).\\
In order to achieve this goal, it is necessary to know when 2 theories have the same correlator functions. The response is given by the following proposition.
\begin{prop}
Two theories possess the same correlator functions when they have the same characteristic function.
\end{prop}
%free action, propagator, and regularization%
Now, we will show why the above renormalization scheme works and leads to the construction of the Wilson effective action.\\
Our first proposition is that the value of a free action depends \emph{additively} on the momentum modes of field components.
\begin{prop} Let M be an Euclidean space, S$_{free}$ a free action on $\cals$(M) given by:
\begin{eqnarray*}
	S_{free}(\phi) := \int_{M^2}d\nu(x)d\nu(y) \,K_{S_{free}}(x,y) \,\phi(x)\phi(y) , \quad K_{S_{free}}(x,y) := \int_{M^*} d\nu(p)\widetilde{K_{S_{free}}}(p)e^{i(p,x-y)},
\end{eqnarray*}
where (K$_{S_{free}}$,$\nu$) is the \emph{kernel distribution} associated to the nondegenerate quadratic form S$_{free}$, \\
then 
\begin{eqnarray}
	S_{free}(\phi) = \int_{M^*} d\nu(p)\, \widetilde{K_{S_{free}^{-1}}}(p)^{-1}\,\widetilde{\phi}(p)\widetilde{\phi}(-p) , 
	\label{acmod}
\end{eqnarray}
where S$_{free}^{-1}$ denotes the inverse form of S$_{free}$, and tilde is the Fourier transform map.
\label{freeex}
\end{prop}
The expression of K$_{S_{free}}$(x,y) is a common feature of the free action in QFT; K$_{S_{free}}$ does not depend on the relative positions x,y but only on the distance between them. \\[10pt]
\textbf{Proof}:\\
One have:
\begin{eqnarray*}
	S_{free}(\phi) &=& \int_{M^2} d\nu(x)d\nu(y)\int_{M^*} d\nu(p)\, e^{i(p,x-y)}\, \widetilde{K_{S_{free}}}(p)\,\phi(x)\phi(y) ,\\
	&=& \int_{M^*} d\nu(p)\,\widetilde{K_{S_{free}}}(p) \int_{M^2} d\nu(x)d\nu(y)\, e^{i(p,x-y)}\,\phi(x)\phi(y),\\ 
	&=& \int_{M^*} d\nu(p)\,\widetilde{K_{S_{free}}}(p) \,\widetilde{\phi}(p)\widetilde{\phi}(-p),
\end{eqnarray*}
so it remains to prove that 
\begin{eqnarray*}
	\widetilde{K_{S_{free}^{-1}}}(p)^{-1} = \widetilde{K_{S_{free}}}(p).
\end{eqnarray*}
By definition, the kernel of the inverse form, called \emph{propagator}, satisfies the relation
\begin{eqnarray*}
	\int_M d\nu(y)\,K_{S_{free}}(x,y)K_{S_{free}^{-1}}(y,z) = \int_M d\nu(y)\, K_{S_{free}^{-1}}(x,y)K_{S_{free}}(y,z) = \delta_z(x) \quad \forall x,z\in M,
\end{eqnarray*}
hence
\begin{eqnarray*}
	&& \int_M d\nu(y) \int_{(M^*)^2} d\nu(p)d\nu(p') \,
e^{i(p,x-y)}e^{i(p',y-z)}\, \widetilde{K_{S_{free}}}(p) \widetilde{K_{S_{free}^{-1}}}(p') = \delta_z(x),\\
  && \int_{(M^*)^2} d\nu(p)d\nu(p')\, \widetilde{K_{S_{free}}}(p) \widetilde{K_{S_{free}^{-1}}}(p')\, e^{i(p,x)}e^{-i(p',z)} \int_M d\nu(y) \,
e^{i(p'-p,y)} = \delta_z(x),\\
  && \int_{(M^*)^2} d\nu(p)d\nu(p')\, \widetilde{K_{S_{free}}}(p) \widetilde{K_{S_{free}^{-1}}}(p')\,e^{i(p,x)}e^{-i(p',z)}\, \delta_{p'}(p) = \delta_z(x),\\
  && \int_{M^*} d\nu(p')\, \widetilde{K_{S_{free}}}(p') \widetilde{K_{S_{free}^{-1}}}(p')\,e^{i(p',x)}e^{-i(p',z)} = \delta_z(x),\\
  && \int_{M^*} d\nu(p')\, \widetilde{K_{S_{free}}}(p') \widetilde{K_{S_{free}^{-1}}}(p')\,e^{i(p',x-z)} = \int_{M^*} d\nu(p) e^{i(p,x-z)}.
\end{eqnarray*}
Q.E.D.\\[10pt]
\textbf{Remark}: The expression (\ref{acmod}) of the free action means that \emph{all} momentum modes of $\phi$ contribute to the value of S$_{free}$($\phi$) because the Fourier transform of the propagator is a nowhere vanishing distribution. In a QFT described at energy scale inferior to $\Lambda$, it is assumed that one want to reduce contributions to Z$_{free}$ which comes from fields whose Fourier transforms are supported on p$^2 \geq \Lambda^2$. There are 2 ways to do this:
\begin{itemize}
	\item the direct approach is to restrict the path integral over the subset of fields whose Fourier transforms are supported on p$^2 \leq \Lambda^2$, 
	\item the second way is to regularize the propagator when the steepest descent theorem applies to QFT.
\end{itemize}

\paragraph{Regularized propagators.} In order to ignore contributions outside I$_{\Lambda'\backslash\Lambda}$ := $\left\{p\in M^*|\Lambda^2 \leq p^2\leq \Lambda'^2\right\}$, we remark that:
\begin{enumerate}[i.]
	\item the above direct method can be done by means of regularization. Indeed, it suffices to replace the Fourier transform of the propagator by the \emph{sharp regularized propagator} $\widetilde{K_{S_{free}^{-1}|\Lambda'\backslash\Lambda}}$ such that: (a) it agrees with $\widetilde{K_{S_{free}^{-1}}}$ on $I_{\Lambda'\backslash\Lambda}$, (b) it becomes 0 outside, and (c) $\lim\limits_{\substack{\Lambda' \to +\infty\\ \Lambda \to 0}} K_{S_{free}^{-1}|I_{\Lambda'\backslash \Lambda}} = K_{S_{free}^{-1}}$. 
\label{shprocond}
	\item from the steepest descent theorem, we may redefine the action in such a manner that this latter is large enough for fields with Fourier modes outside I$_{\Lambda'\backslash\Lambda}$. This can be achieved by modifying the Fourier transform of the propagator into $\widetilde{K_{S_{free}^{-1},\Lambda'\backslash \Lambda}}$ such that: (a) it agrees with $\widetilde{K_{S_{free}^{-1}}}$ on $I_{\Lambda'\backslash\Lambda}$, (b) it decays fastly enough outside I$_{\Lambda'\backslash\Lambda}$, and (c) $\lim\limits_{\substack{\Lambda' \to +\infty\\ \Lambda \to 0}} K_{S_{free}^{-1},I_{\Lambda'\backslash \Lambda}} = K_{S_{free}^{-1}}$.\\
K$_{S_{free}^{-1},\Lambda'\backslash \Lambda}$ will be called a \emph{smooth regularized propagator} with UV cutoff $\Lambda'$ and IR cutoff $\Lambda$. K$_{S_{free}^{-1},\Lambda'\backslash 0}$ and I$_{\Lambda'\backslash 0}$ will be denoted by K$_{S_{free}^{-1},\Lambda'}$ and I$_{\Lambda'}$ respectively.
\label{smprocond}
\end{enumerate}

\textbf{Remarks}: \\
1) We define the regularized propagator in the presence of an IR cutoff because it is requiered by the construction in the next section. \\
2) Here, (regularized) propagators are distributions so it may be possible to obtain relation such as (\ref{acmod}) from a sharp regularized propagator. \\[10pt]
Now, we give some probabilistic proprieties of the regularized free action and its associated Gaussian measure, which will be useful for the next section.

\begin{dfn} Let K$_{S_{free}^{-1};\Lambda}$ be a (smooth or sharp) regularized propagator\footnote{Remark that we use ',' for smooth, '$|$' for sharp, and ';' for arbitrary regularized quantities. }, S$_{free;\Lambda}$ its associated free action (whose kernel is K$_{S_{free}^{-1};\Lambda}$), then its associated \emph{regularized measure} $\mu_{free;\Lambda}$ is defined by: 
\begin{eqnarray*}
	\mu_{free;\Lambda}(f) := \int_{\cals(M)} D\phi \; e^{-S_{free;\Lambda}(\phi)}f(\phi),
\end{eqnarray*}
where f is any formally integrable complex function on $\cals$(M).
\end{dfn}

\begin{prop} For fixed $\Lambda' \geq 0$ and $\Lambda \leq \Lambda'$, we have:
\begin{eqnarray*}
    \mu_{free|\Lambda'} &=& \mu_{free|\Lambda}\ast \mu_{free|\Lambda'\backslash \Lambda}.
\end{eqnarray*}
\label{gausreg}
\end{prop}
\textbf{Proof}:\\
First, let us decompose $\phi \in \cals$(M) into $\varphi,\eta \in \cals$(M) with Fourier transforms supported on I$_{\Lambda}$ and I$_{\Lambda'\backslash\Lambda}$ respectively, an denote by $\cals_{\Lambda'\backslash \Lambda}$(M)$\subset \cals$(M) the subset of fields whose Fourier transforms are supported on I$_{\Lambda'\backslash \Lambda}$. Now, we need the following lemma.
\begin{lem} Let M be a vector space, $\varphi,\eta \in \cals$(M) such that supp($\widetilde{\varphi}$)$\cap$supp($\widetilde{\eta}$) = $\emptyset$,\\
then 
\begin{eqnarray*}
	&& A:= (\widetilde{\varphi} + \widetilde{\eta})(p)(\widetilde{\varphi} + \widetilde{\eta})(-p) = 
  	\begin{cases}
    \widetilde{\varphi}(p)\widetilde{\varphi}(-p) & \text{for } p \in supp(\widetilde{\varphi}) ,\\
     \widetilde{\eta}(p)\widetilde{\eta}(-p) & \text{for } p \in supp(\widetilde{\eta})
\end{cases} \\
	\Leftrightarrow && 
	\left\{ 
\begin{aligned}
  &	(-supp(\widetilde{\varphi})) \cap supp(\widetilde{\eta}) = \emptyset .\\
  & (-supp(\widetilde{\eta})) \cap supp(\widetilde{\varphi}) = \emptyset
\end{aligned}
  \right. 
\end{eqnarray*}
\end{lem}
\textbf{Proof of the lemma}:\\
For $\varphi,\eta \in \cals$(M) such that supp($\widetilde{\varphi}$)$\cap$supp($\widetilde{\eta}$) = $\emptyset$, we have:
\begin{eqnarray*}
  && A = \widetilde{\varphi}(p)\widetilde{\varphi}(-p) + \widetilde{\eta}(p)\widetilde{\varphi}(-p) + \widetilde{\varphi}(p)\widetilde{\eta}(-p) + \widetilde{\eta}(p)\widetilde{\eta}(-p) 
  	= 
  	\begin{cases}
    \widetilde{\varphi}(p)\widetilde{\varphi}(-p) & \text{for } p \in supp(\widetilde{\varphi}),\\
     \widetilde{\eta}(p)\widetilde{\eta}(-p) & \text{for } p \in supp(\widetilde{\eta})
\end{cases} \\
 \Leftrightarrow && \left\{ 
\begin{aligned}
  &	\widetilde{\eta}(-p) = 0 \quad \forall\, p \in supp(\widetilde{\varphi}), \\
  & \widetilde{\varphi}(-p) = 0 \quad \forall\, p \in supp(\widetilde{\eta})
\end{aligned}
  \right. \: \Leftrightarrow \:
  	\left\{ 
\begin{aligned}
  &	(-supp(\widetilde{\varphi})) \cap supp(\widetilde{\eta}) = \emptyset .\\
  & (-supp(\widetilde{\eta})) \cap supp(\widetilde{\varphi}) = \emptyset
\end{aligned}
  \right. 
\end{eqnarray*}
Q.E.D of the lemma.\\[10pt]
From this lemma, we easily obtain:
\begin{eqnarray*}
	S_{free|\Lambda'}(\varphi + \eta) = S_{free|\Lambda}(\varphi) + S_{free|\Lambda'\backslash\Lambda}(\eta).
\end{eqnarray*}
Then, 
\begin{eqnarray}
	&& \int_{\cals(M)} D\phi \; e^{-S_{free|\Lambda}(\phi)} \int_{\cals(M)} D\phi' \; e^{-S_{free|\Lambda'\backslash\Lambda}(\phi')}f(\phi + \phi'), \\
	&=& \int_{\cals_{\Lambda}(M)} D\varphi \; e^{-S_{free|\Lambda}(\varphi)} \int_{\cals_{\Lambda'\backslash \Lambda}(M)} D\eta \;
 e^{-S_{free|\Lambda'\backslash\Lambda}(\eta)}f(\varphi + \eta) 
 \label{shredeq}\\
	&=& \int_{\cals_{\Lambda}(M)} D\varphi \int_{\cals_{\Lambda'\backslash \Lambda}(M)} D\eta \; e^{-S_{free|\Lambda}(\varphi)}\, e^{-S_{free|\Lambda'\backslash\Lambda}(\eta)}f(\varphi + \eta),\\
	&=& \int_{\cals_{\Lambda}(M)} D\varphi \int_{\cals_{\Lambda'\backslash \Lambda}(M)} D\eta \; e^{-S_{free|\Lambda'}(\varphi + \eta)}\; f(\varphi + \eta),\\
	&=& \int_{\cals(M)} D\phi \; e^{-S_{free|\Lambda'}(\phi)}f(\phi).
\end{eqnarray}
The first equality (\ref{shredeq}) comes from the propriety \ref{shprocond}.(b) and the last equality is obtained thanks to the assumption (\ref{mesfico}). Q.E.D.\\[9pt]
%transition sentence%
For smooth regularized measures, equality like (\ref{shredeq}) fails because a rapid decay on interval I of smooth regularized propagator does not guarantee a zero measure for fields supported on I. Nevertheless, one may hope that a rigorous definition of measures on the space of fields will allow to write: 
\begin{ass} For fixed $\Lambda' \geq 0$ and $\Lambda \leq \Lambda'$, we have:
\begin{eqnarray*}
	 \mu_{free,\Lambda'} &=& \mu_{free,\Lambda}\ast \mu_{free,\Lambda'\backslash \Lambda}.
\end{eqnarray*}
\end{ass}
From the above supposition, we can derive the following construction of the Wilson effective action. 

\subsection{Wilson effective action}
By definition, the Wilson effective action allows to describe the low energy regime of a given bare theory by using only degrees of freedom at low energy scales. We will show that it can be obtained by integrating out degrees of freedom, called \emph{fluctuating fields}, defined between the effective and the fundamental scale.\\
Consider a bare theory at scale $\Lambda_0$, given by the (regularized) partition function
\begin{eqnarray*}
	Z_{\Lambda_0}(J) := \int_{\cals(M)} d\mu_{free,\Lambda_0}(\phi) \; e^{-S_{int}(\phi) + \left\langle J,\phi \right\rangle},
\end{eqnarray*}
where $\left\langle J,\phi\right\rangle$ is the image of $\phi$ by J$\in$M*.\\
One want to describe the above theory at smaller energy scale $\Lambda \leq \Lambda_0$, which means in particular that the \emph{exterior} source J is such that $\widetilde{J}(p)$ = 0 for all p$^2 \geq \Lambda^2$ \cite{dagr596}. For this, one defines the \emph{Wilson effective action} S$_{int}^{eff,\Lambda}$ by:
\begin{eqnarray}
	Z(\Lambda,J) := \int_{\cals(M)} d\mu_{free,\Lambda}(\phi)\; e^{- S_{int}^{eff,\Lambda}(\phi) + \left\langle J,\phi \right\rangle} = Z_{\Lambda_0}(J) \quad \forall \Lambda \leq \Lambda_0.
	\label{weq}
\end{eqnarray}
From this definition, theories (S$_{int}$,$\Lambda_0$) and S$_{int}^{eff,\Lambda}$ have identical correlation functions, then S$_{int}^{eff,\Lambda}$ is an effective theory of (S$_{int}$,$\Lambda_0$) at scale $\Lambda$. Elements of  $\cals_{I_{\Lambda_0}}$(M) can be called \emph{regularized fields}. \\[10pt]
To construct the Wilson action, one decompose a regularized field $\phi$ into sum of fields $\varphi$ and $\eta$, with momentum modes supported on I$_{\Lambda}$ and  I$_{\Lambda_0\backslash\Lambda}$ respectively \footnote{$\varphi$ and $\eta$ must agree on their common boundary.}. Using Proposition \ref{gausreg}, one obtain:
\begin{eqnarray*}
	Z_{\Lambda_0}(J) &=& \int_{\cals(M)} d(\mu_{free,\Lambda} \ast \mu_{free,\Lambda_0\backslash\Lambda})(\varphi + \eta)\, e^{- S_{int}(\varphi + \eta) + \left\langle J,\varphi + \eta \right\rangle}, \\
	&=& \int_{\cals(M)} d\mu_{free,\Lambda}(\varphi) \int_{\cals(M)}  d\mu_{free,\Lambda_0\backslash\Lambda}(\eta)\, e^{- S_{int}(\varphi + \eta) + \left\langle J,\varphi \right\rangle}.
\end{eqnarray*}
So we deduce the Wilson effective action from the equality
\begin{eqnarray*}
	e^{- S_{int}^{eff,\Lambda}(\varphi)} = \int_{\cals(M)} d\mu_{free,\Lambda_0\backslash\Lambda}(\eta) e^{- S_{int}(\varphi + \eta)}.
\end{eqnarray*}
In the above construction, $\varphi$ is called the \emph{background field} and $\eta$ is the \emph{fluctuating field} of the effective theory.\\[10pt]
The last step of the renormalization theory is the renormalization procedure which consists to remove the UV cutoff to infinity which will not treated in the present paper. 
%mila ampiana%
We will end this review on physical concepts of renormalization theory with a probabilistic derivation of the Legendre effective action.
 
\subsection{Legendre effective action}
The last construction which involves convolution products in QFT we shall explore is those of the Legendre effective action $\Gamma_S$. For a QFT with action S, we can (formally) associate the measure $\mu_S$ in the same manner as for the free action case. Following lectures of Krzysztof Gaw\c edzski, we will show that $\Gamma_S$ arises in a large N convolution product of $\mu_S$ \cite{krga96}.\\[10pt]
Therefore, let us consider the measure $\mu_S$ defined by
\begin{eqnarray*}
	\mu_S(f) := \int_{\cals(M)}D\phi \; e^{-S(\phi)}\, f(\phi),
\end{eqnarray*}
for any formally integrable $\C$-valued function f on $\cals$(M), and define the empirical N-mean map $\Sigma_N$ by: 
\begin{eqnarray*}
	\Sigma_N : \oplus^N \cals(M) \ni \: \bar{\xi} := (\xi_j)_{1\leq j \leq N} \mapsto \frac{1}{N}\sum_{j = 1}^N\xi_j \in \cals(M).
\end{eqnarray*}
By using the map $\Sigma_N$, one can (formally) define the following convolution product of measures:
\begin{eqnarray*}
	 \int_{\cals(M)} d(\underbrace{\mu_S\,\ast_{L}...\ast_L\, \mu_S}_{\textrm{N times}}) (\phi)\; f(\phi) &:=& \int_{\cals(M)} d\mu_S(\xi_1)\, ... \int_{\cals(M)} d\mu_S(\xi_N) \; f(\frac{1}{N}(\xi_1 + ... + \xi_N)),\\
 \textrm{or }\quad \underbrace{\mu_S\,\ast_L...\ast_L\, \mu_S}_{\textrm{N times}} &:=& \mu_S^N \circ C(\Sigma_N), 
\end{eqnarray*}
where $\mu_S^N$ is the N-th measure product of $\mu_S$ by himself, and C($\Sigma_N$) the pullback of $\Sigma_N$. One easily remarks that $\ast_L$ is not associative.\\
Let us denote $\underbrace{\mu_S\,\ast_L...\ast_L\, \mu_S}_{\textrm{N times}}$ by $\mu_{\Gamma,N}$ and express the measure $\mu_{\Gamma,N}$ in the form
\begin{eqnarray*}
	d\mu_{\Gamma,N}(\phi) =: e^{-N\tilde{\Gamma}(\phi)}D\phi. 
\end{eqnarray*}
Using the notation e$^{W(J)}$ := $\int_{\cals(M)}d\mu_S\, e^{<J,\phi>}$ for J $\in\cals^*$(M), a rough calculation of $\tilde{\Gamma}(\phi)$ gives
\begin{eqnarray*}
	e^{-N\tilde{\Gamma}(\phi)} = \int \delta(N\phi - \phi') e^{-N\tilde{\Gamma}(\phi'/N)}d\mu_L &=& \int \delta(N\phi - N(\phi'/N))dP_N(\phi'/N),\\
	&=& \int \delta(N\phi - N\Sigma_N(\bar{\xi}))\prod_{j = 1}^Nd\mu_S(\bar{\xi}),\\
	&=& \int \int DJ e^{-<N\phi - \sum_{j=1}^N\xi_j,J>} \prod_{j = 1}^Nd\mu_S(\xi_j),\\ 
	&=& \int DJ \int \prod_{j= 1}^N d\mu_S(\xi_j) e^{-<N \phi - \sum_{j=1}^N\xi_j,J>},\\
	&=& \int DJ e^{-N<\phi,J> + NW(J)}.
\end{eqnarray*}
For large N and by supposing that the steepest descent theorem applies, we obtain :  
\begin{eqnarray*}
		e^{-N\tilde{\Gamma}(\zeta)} = e^{- sup_{J\in \cals^*(M)} \left\{<\zeta,J> - W(J)\right\}N + o(N)}
	=: e^{-N\Gamma(\zeta) + o(N)}.
\end{eqnarray*}
The quantity 
\begin{eqnarray}
	\Gamma(\zeta) := sup_{J\in \cals^*(M)}\left\{<\zeta,J> - W(J)\right\} 
\end{eqnarray}
is called the \emph{Legendre effective action} of the theory.\\[10pt]
\textbf{Remark}: A well-known important fact on Legendre effective action is that the expectation value $\left\langle ev_x\right\rangle$, i.e. the 'average' field, is a critical point of the Legendre effective action. For a QFT with classical action S, one can deduce that the Legendre effective action $\Gamma_S$ is a classical effective theory of S in the sense that its (classical) Euler-Lagrange equation admits a solution of quantum nature. \\[10pt]
%mila faranana%	
Following these different constructions in renormalization theory, one concludes that a given physical theory depends on a given characteristic scale which, in fact, is fixed by physical measurements. In the following section, we will propose a mathematical framework where one can deal with scale dependent theories. 

\section{Scale dependence in C*-algebraic models}
In this section, we will propose a mathematical structure which reflects naturally the scale dependence of physical theories; this is the C*-algebraic state space structure and its hierarchy. \\
An amazing feature of this hierarchy is that spaces of higher scales are built from those of smaller scales. This may be indicate in part how to build physical theories from smaller energy scales. \\
As seen the above section, an effective theory of some fundamental theory A can be obtained by 'rearranging' fundamental degrees of freedom in such a way so one can reduce them into few degrees of freedom enable to describe A at low energy. In other words, the state space of a fundamental theory is much larger than those of its effective theory. Our hierarchy will also possess this physical propriety. So let us begin the construction of the hierarchy.
\\[10pt]
From a compact Hausdorff space X (a space-time), we will construct a hierarchy which will be used to be the state space of classical systems on X. This consists in building a topological structure on the set SC$^n$(X) := $\underbrace{S(C(...S(C}_{n \; times}$(X) for n$\in \N$, where SC(X) is the set of positive normalized linear forms on the C*-algebra C(X) of continuous complex functions on X. \\
Firstly, we will consider the relative weak*-topology on SC(X), for which this latter is Hausdorff compact \cite{hera04,npla98}, to generate a topology on our hierarchy. In this case, one notice that the construction of the weak*-topology for higher scales of the hierarchy is natural.\\
One can define SC$^n$(X) in a recursive way; it suffices to build \emph{in the same time} a generalization of the Gelfand transform tg and a generalization of the relative weak*-topology.
\begin{dfn} Let $\tg$ be a map defined, at scale i$\in N$, by:
\begin{eqnarray*}
	\tg_i : C(SC^{i-1}(X)) &\rightarrow& Map(SC^i(X);\C), \quad i\in \N^* \\
	       f &\mapsto& \tg_i(f)(\mu) := \mu(f) \quad \forall \mu \in SC^i(X).
\end{eqnarray*}
where SC$^{i-1}$(X) is equipped with the topology $\T^*_{i-1}$ generated by $\left\{\tg_{i-1}(f)^{-1}(O) | f \in C(SC^{i-2}(X)), O \in T_{\C}\right\}$,
then $\tg$ and $\T$* are called \emph{generalized Gelfand transform} and  \emph{generalized weak*-topology} respectively.
\end{dfn}

Due to the recursive construction of $\tg$, the topology $\T$* is Hausdorff compact. Consequently, one have the inclusion $\delta$ defined by\footnote{The injectivity of $\delta$ follows from Urysohn's lemma which says that C(X) separates points on X for Hausdorf compact X.}:
\begin{eqnarray*}
	\delta^{i+1}: SC^i(X) &\rightarrow& SC^{i+1}(X),\\
	         w &\mapsto& \delta^{i+1}_w(f) := f(w) \quad \forall f \in C(SC^i(X)). 
\end{eqnarray*}

\begin{prop} Let $|_i$ be the restriction map to SC$^i$(X), then the relative topology T*$_i|_{i-1}$ on SC$^{i-1}$(X) is homeomorphic to T*$_{i-1}$.
\end{prop}
Indeed, on the one hand, from the relation \textbf{tg}$_i$(f)$\circ \delta^i$ = f, for all f$\in$C(SC$^{i-1}$(X),T*$_{i-1}$), one easily deduce that f is continuous on T*$_i|_{i-1}$ and that T*$_{i-1} \prec$ T*$_i|_{i-1}$ by construction of T*. \\
On the other hand, by noticing that the compactness and 'Hausdorff-ness' are hereditary \footnote{A topological propriety on a space X is \emph{hereditary} whenever it is also possessed by any subspace of X.} and that for topologies T and T', Hausdorff and compact respectively, such that T is weaker than T', T and T' must coincides, then one achieves the proof.\\[10pt]
Now, one can define the hierarchy (SC*(X),$\T$*) \footnote{We will denote (SC*(X),$\T$*) by SC*(X) when there is no risk of ambiguity.} as being the filtration $\delta^{i-1}$: (SC$^{i-1}$(X),$\T^*_{i-1}$) $\hookrightarrow$ (SC$^i$(X),$\T^*_i$). An observable on SC*(X) is then a collection of observables defined on each scale of SC*(X) which satisfy some compatibility relations.

\begin{dfn} Let F := $\left\{F_i\in C(SC^i(X)|i\in \N\right\}$ such that:
\begin{eqnarray*}
	F_i|_k = F_j|_k \quad \forall i\geq k, j\geq k,
\end{eqnarray*}
then F is called an \emph{observable} on SC*(X).
\end{dfn}

In practice, an observable on SC*(X) is determined by its restrictions on each scale of SC*(X). In fact, a restriction of F describes its behavior on a given scale. Moreover, thanks to the filtration map $\delta$, any restriction F$|_i$ determines completely F$|_j$ for j$\leq$i. However, one can use the Gelfand transform to move up scales.  \\[10pt]
\textbf{Remarks}:\\
1) For X = $\left\{point\right\}$, the hierarchy SC*($\left\{point\right\}$) is given by: SC$^n$($\left\{point\right\}$) = $\left\{point\right\}$ $\forall n\in \N$. In other words, theories on $\left\{point\right\}$ are trivially identical at any scale. 
%mbola tsy ampy%

\subsection{Classical fields on SC*(X)}
In order to explain scale dependence in QFT in mathematical terms, we begin with a mathematical framework for the classical field theory: the \emph{fiber bundle theory}. \\
Let us recall that for Hausdorff compact topological spaces E,X,F, a \emph{fiber bundle} p:E$\rightarrow$X with \emph{typical fiber} F is a continuous surjection such that E is locally homeomorphic to X$\times$F. A \emph{global section} on the fiber bundle E is a continuous map which sends x$\in$ X to a point of the fiber  p$^{-1}$(x). \\
Our choice is to see sections of a vector bundle as representing classical fields. Therefore, the set $\Gamma$(E) of global sections on E can be interpreted as a semi-quantum state space because it is a representation space of C(X). On the one hand, it have a classical propriety because the observable algebra is commutative. \\
To construct classical fields on the hierarchy SC*(X), it is more convenient to use the algebraic dual of a n-dimensional vector bundle over X which is an idempotent of End(C(X)$^n$) according to the Serre-Swan theorem. So let p be an idempotent on C(SC$^{i}$(X))$^n$, then one wants to build an idempotent on C(SC$^{i+1}$(X))$^n$ from p. 
\begin{prop} Let p be an idempotent on C(SC$^{i}$(X))$^n$, then the map
\begin{eqnarray*}
	\textbf{tg}_{i+1}\circ p \circ C(\delta^{i+1}): C(SC^{i+1}(X))^n \rightarrow C(SC^{i+1}(X))^n,
\end{eqnarray*}
is an idempotent on C(SC$^{i+1}$(X))$^n$. 
\end{prop}
\textbf{Proof}:\\
It suffices to prove that:
\begin{eqnarray*}
	C(\delta^{i+1}) \circ \tg_{i+1} = Id_{i+1}, 
\end{eqnarray*}
where Id$_{i+1}$ is the identity map on C(SC$^{i+1}$(X))$^n$. \\
One have:
\begin{eqnarray*}
	C(\delta^{i+1}) \circ \tg_{i+1}(f) &=& C(\delta^{i+1})(\tg_{i+1}(f)) \quad \forall f\in C(SC^{i+1}(X))^n,\\
	&=& \tg_{i+1}(f)\circ \delta^{i+1},\\
	&=& f.
\end{eqnarray*}
Q.E.D.\\[10pt]
From the above proposition, it follows that a classical field p$\in$Idem(C(X)$^n$)  allows to define a \emph{generalized classical field} \textbf{tg}$_{i}\circ$ p $\circ$ C($\delta^{i}$)$\in$Idem(C(SC$^{i}$(X))$^n$).\\[10pt]
Now, we will show that interaction can be introduced via convolution product of the free measure with a term which will play the the interacting term.  

\section{Sequence Construction of interaction}

In this section, we will develop some constructions which may enhance the measure-theoretic approach to quantum field theory.
As seen in the first section, one assumes the existence of a Gaussian measure $\mu_{free}$ in the space of fields, and interacting theories are obtained by namely adding a supplementary (interacting) term to the free action. However, this last step conducts to divergence problems. Our idea is to introduce the interacting term by means of convolution operation as done in some constructions in probability theory when one deals with sequences of dependent random variables.

\subsection{Interacting sequences}

In probability theory, theorems on the weak convergence to a normal law, such as the Lindeberg-Feller theorem \cite{ribap01}, works essentially for sequences of independent random variables. More precisely, one consider a sequence of independent random variables and then its partial sum process; under some additional conditions on mean and variance of the initial sequence, the partial sum process converges weakly to a normal random variable. These conditions on mean and variance of the  sequence are not so important in the sense that they do not depend on the values of these two quantities. Roughly speaking, the partial sum process of a sequence of independent random variables is inclined to follow a normal law.
\par
On the other hand, free physical systems such as free QFTs are often described by a quadratic action, i.e. by normal laws in the path integral formalism. sequence. Therefore, one may suggest:
\begin{ass} A free physical system can be represented by the partial sum process of an independent random variables sequence. More generally, an interacting physical system can be represented by sequence of dependent random variables. 
\end{ass}
It is well-known that the probability law of a sum of independent (not necessarily equally distributed) random variables is given by the convolution product of random variable's laws. One deduce from the above explanation that a sequence of convolutions of probability laws converges weakly to a normal law when its mean and variance satisfy some technical conditions. \\ 
Now, we will show that the probability law of an interacting sequence can also be obtained by a convolution product of its free probability law. To illustrate this affirmation, we will examinate two simple cases.

\paragraph{A. Interactions on Bernouilli sequences.}
A sequence of Bernouilli random variables (Be$_i$), i$\in \N$, allows to define a generalized binomial random variable Bi$_n$ := $\displaystyle\sum_{i\leq n} Be_i$ which is, by definition, the partial sum of order n of (Be$_i$). Intuitively, the binomial random variable is given by some succession of Be$_i$ tests. When (Be$_i$) is free, then its partial sum process of order n gives the usual binomial Bi$_{n,free}$ of order n which follows the law p$_{n,free}$ given by:
\begin{eqnarray*}
	p_{n,free}(k) := p_{Be_1}*...*p_{Be_n}(k), \quad k\in \N,
\end{eqnarray*}
where the associative convolution product is defined by: 
\begin{eqnarray}
	f*g := m_{\C}\circ (f\tp g)\circ\Delta^+, \quad \Delta^+(k) := \sum_{\substack{a+b = k\\a,b\in \N}}a\oplus b, \quad k\in \N, \quad f,g\in Map(\N,\C).
	\label{stacovpr}
\end{eqnarray}
and m$_{\C}$ is the multiplication map on $\C$.\\
When (Be$_i$) is interacting, then we propose 2 ways to introduce interacting terms of Bi$_n$'s law.
\begin{enumerate}
	\item The first way is to introduce interacting term by pointwise product with the free probability. For probabilities having densities, one may consider the pointwise product of the interacting term by the free probability density. For the binomial case, we define a probability law p$_n$ of an interacting sequence by:
\begin{eqnarray}
	p_n = p_{n,free}.p_{int},
\end{eqnarray}
where p$_{int}$ is a real function such that:
\begin{eqnarray*}
	0 \leq p_{n,free}.p_{int} \leq 1 \quad \textrm{and} \quad
	\sum_{k=0}^n p_{n,free}(k)p_{int}(k) = 1. 
\end{eqnarray*}
In this setting, the construction of the interacting term amounts to find a random variable with law p$_{n,free}$ and mean one. Transformations $p_{n,free} \rightarrow p_{n,free}.p_{int}$ , f $\rightarrow$ f, are equivalent to $p_{n,free} \rightarrow p_{n,free}$ , f $\rightarrow p_{int}$.f . \\
Let us show an explicit construction of such interaction on a Bernouilli sequence with parameter p, 0 $\leq$ p $\leq$ 1. The interacting term is built from the relation:
\begin{eqnarray*}
	\sum_{k=0}^n C^k_np^k(1-p)^{n-k} = (p +(1-p))^n = 1.
\end{eqnarray*}
Now, one may perturb coefficients p and (1-p) by 2 positive reals a,b\footnote{a,b may depend on n.}, then one obtains:
\begin{eqnarray}
	&& \sum_{k=0}^n C^k_n(ap)^k(b(1-p))^{n-k} = (ap + b(1-p))^n ,\\
\textrm{i.e.} && \sum_{k=0}^n C^k_np^k(1-p)^{n-k}\left[a^kb^{n-k}(ap + b(1-p))^{-n}\right] = 1.	\label{nint}
\end{eqnarray}
\begin{prop} The expression $a^kb^{n-k}(ap + b(1-p))^{-n}$ is an interacting term.
\end{prop}
\textbf{Proof}:\\
From the equality (\ref{nint}), it remains to prove that:
\begin{eqnarray*}
	0 \leq C^k_np^k(1-p)^{n-k}\left[a^kb^{n-k}(ap + b(1-p))^{-n}\right] \leq 1 \quad \forall k = 0,1,...n.
\end{eqnarray*}
In (\ref{nint}), one have a sum of n positive nonzero terms which gives 1; therefore, if one of these terms is superior to 1, then the sum would be superior to 1.\\
Q.E.D.\\ 
Some particular cases:
\begin{enumerate}
	\item When a) a.p + b(1-p) = 1, and b) a,b does not depend on n, then the law p$_{n,free}(p).p_{n,int}$, with	p$_{n,int}$(k) = a$^k$b$^{n-k}$, is identical to the law p$_{n,free}$(a.p) of parameter a.p.\\
In this case, the interacting term a$^k$b$^{n-k}$ allows to pass from a free theory with parameter p to another one with parameter a.p.
	\item When a) a.p + b(1-p) = 1, and b) $\lim\limits_{n\to \infty}$n.a.p = $\lambda$, then the law p$_{n,free}(p).p_{n,int}$, with p$_{n,int}$ = a$^k$b$^{n-k}$, converges weakly (following n) to the Poisson law with parameter $\lambda$.
\end{enumerate}
  \item The second way is to multiplicate interacting term with the free term by means of a convolution product. For probabilities having densities, we will consider the convolution of the free probability density with the interacting term. For our binomial law, we define a probability law \^p$_n$ of an interacting sequence by:
\begin{eqnarray*}
	\hat{p}_n = p_{n,free}*\hat{p}_{int},
\end{eqnarray*}
with analogous conditions to those of the first construction, i.e. 
\begin{eqnarray}
	0 \leq p_{n,free}*\hat{p}_{int} \leq 1 \quad \textrm{and} \quad \sum_{k\in \N} p_{n,free}*\hat{p}_{int}(k) = 1.
	\label{codcovpr}
\end{eqnarray}
Our first remark is that \^p$_{int}$ is necessarily a probability law when one uses the convolution product defined in (\ref{stacovpr}); in particular, \^p may be a free binomial law (with order higher than n)\footnote{It explains why the summation in (\ref{codcovpr}) should be taken on $\N$.}. One obtains the same result for probabilities having densities when one uses the standard convolution product on L$^1$(\textbf{R}).
\end{enumerate}

\textbf{More on the convolution construction of the interaction}: In our setting, the main feature of the convolution product's use is that the probability \^p$_n$(k), k$\in \N$, depends on p$_{n,free}$(j) for some j$\in \N$ near k. From this propriety, we will give the following interpretation of the interaction. \\
First, we begin with few definitions.
\begin{dfn} For a discrete probability p on \textbf{R}, i.e. $\sum_{i\in J}$p(i) = 1 for finite set J$\subset$\textbf{N}, let m$\in$\textbf{N} such that p(m) $\neq$ 0 and p(k) = 0 $\forall$ k $>$ m, then m =: Order(p) will be called the \emph{order} of p, and $\left\{i\in \N\,|\, i\leq Order(p) \right\}$ =: Conf(p) is the \emph{configuration space} of p.
\end{dfn}

Now, when one consider an interacting term \^p$_{int}$ such that Order(\^p$_{int}$) $\leq$ Order(p$_{free}$), then there exists an \emph{injective} correspondance $\Xi$ from Conf(p$_{free}$$\ast$\^p$_{int}$) to the collection of subsets of Conf(p$_{free}$), i.e. the hierarchy configuration space on Conf(p$_{free}$) according to our precedent section. Therefore, the presence of such interaction \^p$_{int}$ decribes subspace in the Conf(p$_{free}$)'s hierarchy whose scale is characterized by the quantity Range(\^p$_{int}$) := Order(\^p$_{int}$) + 1. Indeed, for k$\in$Conf(p$_{free}\ast$\^p$_{int}$), we have: card($\Xi$(k)) $\leq$ Range(\^p$_{int}$). \\[10pt]
\textbf{Proprieties}:
\begin{enumerate}
	\item No interaction: when Range(\^p$_{int}$) = 1, then \^p$_{int}$(0) = 1 and p$_{free}\ast$\^p$_{int}$ = p$_{free}$. In other words, the interaction does not exist when its range or the number of interacting neighbours is 1. In particular, the configuration space does not change.
	\item Interacting configuration space: when Range $\geq$ 2, i.e. when there is an interaction, then the interacting configuration space Conf$_{free*int}$ is bigger than the old one \emph{but} does not contained this latter. Is it the case in physical theories? 
	\item Representation of intracting states: let \^p$_{int}$ be an interaction of order r, and p$_{free}$ a free theory on a configuration space Conf$_{free}$ such that card(Conf$_{free}$) = n. A thorough analysis of Conf$_{free*int}$ leads to the representation of interacting states by (r+1)-tuplet of nondecreasing successive free states, classified into 3 types : 
\begin{itemize}
	\item Conf$_{free*int}\ni$a+r $\sim$ (a,a+1,...,a + r) for 0 $\leq$ a $\leq$ n - r, 
	\item Conf$_{free*int}\ni$r-a $\sim$ ($\underbrace{0,...,0}_{\textrm{a times}}$,0,1,2,...,r - a) for 1$\leq$ a $\leq$ r, 
	\item Conf$_{free*int}\ni$b+n-r $\sim$ (b + n - r,...,n - 1,n,$\underbrace{n,...,n}_{\textrm{b times}}$) for 1$\leq$ b $\leq$ r,
\end{itemize}
The 2 last types concern states which contain self-interacting 'pure states'.
	\item Trivial interaction: when Range(\^p$_{int}$) $\geq$ 2 and \^p$_{int}$(Order(\^p$_{int}$)) = 1, then the interacting state behaves like free state in the sense that for each state j of Conf$_{free*int}$ corresponds an unique state i such that p$_{free}\ast$\^p$_{int}$(j) = p$_{free}$(i).
\end{enumerate}
The main feature of interacting theory for probabilities having densities is that they are non local, \footnote{A theory is non local when the interaction at a given point x depends on points separated by a finite distance to x.} when the range of the interaction is finite. In order to obtain local interacting theory, the range of the interaction should be infinitely short; it can be obtained with the use of derivatives.
\begin{enumerate}
	\item No interaction: When dens$_{int}$ = $\delta$, then dens$_{free}\ast$dens$_{int}$ = dens$_{free}$. Remark that the support of the Dirac distribution is a singleton, therefore each point of Conf$_{free}$ have no interacting neighbour. In the continuous case, the support K of the interacting term determines the family of interacting neighbours; we said that the range of the interaction is finite when the support K is an 'usual' subset of $\R$.  
	\item Representation of interacting states: let dens$_{int}$ be an interacting term with support K$\subset\R$, and dens$_{free}$ be the density probability of a free theory on $\R$, then interacting states can be identified with fibers at free states of a vector bundle or with maps
\begin{eqnarray*}
	s: \R &\rightarrow& Subset(\R),\\
	    x &\mapsto& x - K.
\end{eqnarray*}
\end{enumerate}
It is not difficult to derive a formal analogy of the above construction for Quantum Field Theory. However, the real value of resulted theory should be checked by more calculations. 
\paragraph{B. Gauge theory.} 
In Particle Physics, forces and interactions are explained by gauge theory. Its main feature is that a free Lagrangian is not invariant under some local transformations on matter fields unless one introduces a supplementary term containing a new 'field', the \emph{gauge potential}, which mediates interaction between matter fields.\\
The sequence construction of interaction applied to QFT conducts to the following consideration. Let Fields$_m$ be a vector space of matter fields, Pot$_g$ a vector  space of 'gauge' potentials\footnote{A priori, the term 'gauge' is not adequate because the present construction does not involve any concept of gauge theory.}, B$_m$ and B$_g$ 2 nondegenerate bilinear forms (free actions) on Fields$_m$ and Pot$_g$ respectively. Let us define the formal partition function of an interaction $\Sigma$ between Fields$_m$ and Pot$_g$ by:
\begin{eqnarray}
 Z_{free*int} := \int_{Fields_m} D\phi \int_{Pot_g} DA\; e^{-\frac{1}{2}B_m(\Sigma(A)\phi,\Sigma(A)\phi)}e^{-\frac{1}{2}B_g(A,A)},
	\label{scpart}
\end{eqnarray}
where $\Sigma$ is a linear map from Pot$_g$ to End(Fields$_m$) such that the bilinear form  B$_{m,\Sigma(A)}$ := B$_m$($\Sigma(A)\cdot, \Sigma(A)\cdot$) is nondegenerate for any gauge potential A and det(B$_{m,\Sigma(A)}$) does not depend on A.\\
With these two conditions, one remarks that Z$_{free*int}$ is easily normalized by multiplicative factor as in the free case.

\section{Conclusion}
We have seen some insights of probability theory in the formulation of QFT within the path integral formalism. In addition, the uses of convolution products in the Kreimer-Connes approach of perturbative renormalization and the effective theory of non perturbative renormalization have lead us to a probabilistic construction of interacting theories, modestly developed on simple models such as Bernouilli sequences. Our construction is devoted to work in QFT because in our setting the partition function is naturally normalized for free as well as interacting theories. However, it remains formal because constructed in the realm of path integral formalism.\\
Our future work will be concerned with further development of the sequence construction of interaction for Quantum Field Theory.


\begin{thebibliography}{10}
\bibitem{dagr196} David Gross, \emph{Renormalization Group: Lecture 1}, QFT Program at IAS, \textbf{1996}.
\bibitem{dagr596} David Gross, \emph{Renormalization Group: Lecture 5}, QFT Program at IAS, \textbf{1996}.
\bibitem{edwi96} Edward Witten, \emph{Perturbative renormalization: Lecture 1}, QFT Program at IAS, \textbf{1996}.
\bibitem{krga96} Krzysztof Gaw\c edzski, \emph{Conformal Field Theory}, QFT Program at IAS, \textbf{1996}.
\bibitem{paet02} Pavel Etingof, \emph{Mathematical ideas and notions of Quantum Field Theory}, OpenCourseWare at MIT, \textbf{2002}.
\bibitem{frwi98} Franck Wilczek, \emph{Quantum Field Theory}, preprint: hep-th/9803075, \textbf{1998}.
\bibitem{acdk99} Alain Connes, Dirk Kreimer, \emph{Renormalization in Quantum Field Theory and the Riemann-Hilbert problem I: the Hopf algebra structure of graphs and the main theorem}, preprint: hep-th/9912092, \textbf{1999}.
\bibitem{acdk00} Alain Connes, Dirk Kreimer, \emph{Renormalization in Quantum Field Theory and the Riemann-Hilbert problem II: the $\beta$-function, diffeomorphisms and the renormalization group}, preprint: hep-th/0003188, \textbf{2000}.
\bibitem{hijo04} H\'ector Figueroa and Jos\'e F. Gracia-Bondia, \emph{The uses of Connes and Kreimer's algebraic formulation of renormalization theory}, preprint: hep-th/0301015, \textbf{2004}.
\bibitem{viri02} Vincent Rivasseau, \emph{An introduction to Renormalization}, S\'eminaire Poincar\'e, \textbf{2002}.
\bibitem{doma02} Dominique Manchon, \emph{Hopf algebras, from basics to applications to renormalizations}, Extended version of lectures given at Bogota University, \textbf{2002}.
\bibitem{hera061} Herintsitohaina M. Ratsimbarison, \emph{Feynman diagrams, Hopf algebras and renormalization}, preprint: hep-th/, \textbf{2006}.
\bibitem{hera04} Herintsitohaina M. Ratsimbarison, \emph{Dualit\'e non-commutativa de Gelfand-Naimark et applications en Th\'eorie de Jauge et Structure de spin$^c$}, M\'emoire de DEA, Facult\'e des Sciences d'Antananarivo, \textbf{2004}.
\bibitem{jadi78} Jacques Dixmier, \emph{Alg\`ebres d'Op\'erateurs}, Rendiconti S.I.F. - XLV.
\bibitem{ribam98} Richard F. Bass, \emph{A brief introduction to measure theory and Integration}, On-line course notes, University of Connecticut, \textbf{1998}.
\bibitem{ribap01} Richard F. Bass, \emph{Probability Theory}, On-line course notes, University of Connecticut, \textbf{2001}.
\bibitem{npla98} Nicolas P. Landsman, \emph{Lecture notes on C*-algebras, Hilbert C-modules and quantum mechanics}, preprint: math-ph/9807030, \textbf{2003}.
\bibitem{npla03} Nicolas P. Landsman, \emph{Lecture notes on C*-algebras and K-theory}, N.P. Lansdman's homepage, draft: \textbf{2003}.
\end{thebibliography}
\end{document}